\documentclass[12pt]{article}
\usepackage{a4}
\usepackage{epsfig,graphicx}

\newcommand\ignore[1]{}
\begin{document}

\hfill To be published in Proceedings of ECRYS-99,

\hfill Journ. de Physique, Coll., December 1999.

\vskip .6in

\begin{center}
X-RAY DIFFRACTION\ FROM\ PINNED\ CHARGE\ DENSITY\ WAVES
\end{center}

\centerline{S. Rouzi\`{e}re$^{1,3}$, S. Ravy$^{1}$, S. Brazovskii$^{2}$
and J.-P. Pouget$^{1}$} \vspace{1cm}

\noindent \textit{{$^{1}$Laboratoire de Physique des Solides (CNRS UMR
8502), B\^{a}t. 510,\newline
$^{\ }$Universit\'{e} Paris-sud, 91405 Orsay C\'{e}dex, France}}

\noindent\textit{{\ $^{2}$Laboratoire de Physique Th\'{e}orique et
Mod\`{e}les Statistiques (CNRS\newline
$^{\ }$UMR 8626), B\^{a}t.100, Universit\'{e} Paris-Sud, 91406 Orsay
C\'{e}dex, France}}

\noindent \textit{{$^{3}$Current address : Institute for Solid State
Physics, University of Tokyo,\newline
$^{\ }$Japan}} \parskip 6pt \vskip .2in

\begin{narrow}
  {\small
  {\bf Abstract.}
We present an x-ray study of doped charge density waves systems. When
a $2k_{f}$-charge density wave is strongly pinned to impurities, an interference
effect gives rise to an asymmetry between the intensities of the $+2k_{f}$ and
$-2k_{f}$ satellite reflections. Moreover, profile asymmetry of the satellite reflections
 indicates the existence of Friedel oscillations (FOs) around the defects.
We have studied these effects in  V- and W-doped blue bronzes.
A syncrotron radiation study of the V-doped blue bronze clearly reveals
the presence of FO around the V atoms.
}
 \end{narrow}

\vskip .2in \textbf{1. INTRODUCTION} \vskip .1in The pinning of charge
density waves (CDW) on impurities has a crucial influence on the collective
properties of quasi-one-dimensional (1D) metals [1]. Although the
interaction between CDW and impurities has been studied for more than two
decades in quasi-1D systems, it is far from being correctly understood. In
particular, the presence of Friedel oscillations (FOs) around the defect and
their influence on the CDW order, though suggested theoretically [2], has
not received experimental confirmation so far. Recently, a new x-ray
diffraction effect resulting from the interference between the CDW
displacive contribution and the disorder contribution has been able to bring
some valuable information and a thorough insight into the local description
of the CDW-defect interaction [3]. With this technique, the phase of the CDW
at the impurity position together with local distortions of the CDW phase
around the impurity become measurable.

The theory of the scattering of CDW pinned on impurity has already been
presented [3] and the main results can be summarized as followed. The $%
2k_{f} $-periodic lattice distortion (PLD) in quadrature with the $2k_{f}$%
-CDW, gives rise to satellite reflections at the reciprocal positions $\pm
2k_{f}$ around each Bragg reflection. The main point is that the spatial
coherence between the impurity position and the phase of the PLD (the strong
pinning case) can result in interferences between the satellite reflections
and the diffuse scattering due to the chemical disorder. At first-order,
this effect gives rise to a $+2k_{f}/-2k_{f}$ intensity asymmetry (IA) of
the satellite reflection. The direction of this asymmetry depends on the
pinning value of the phase ($\Phi _{0}$) of the PLD on each impurity site.
It allows to determine whether the first-neighbor atoms are displaced away ($%
\Phi _{0}=0$) or towards ($\Phi _{0}=\pi $) the impurity. Moreover, in
presence of a change of the wave length of the modulation around the
impurity, a profile asymmetry (PA) of each satellite reflection is expected.
This is the case of FOs which will be discussed later. These asymmetry
effects have been first discovered in the organic charge transfer salts of
the TTF-TCNQ family [3,4], giving rise to ''white lines'' on the diffuse
scattering patterns. Here, we report an X-ray study of vanadium-(2.8\%) and
tungsten-(2\%) doped blue bronzes.

Blue bronze K$_{0.3}$MoO$_{3}$ crystallizes in the monoclinic $C$2/m spatial
group with the cell parameters a=18.25\AA , b=7.56\AA , c=9.885\AA\ and $%
\beta $=117.5$^{o}$. MoO$_{6}$ octahedra stack along the $\mathbf{b}$
direction and form layers along the [102] direction. Below the Peierls
transition at $T_{p}$=183K [5], the PLD associated to the charge density
wave gives rise to satellite reflections located at the reciprocal reduced
wavevector $\mathbf{q}_{c}=(1,2k_{F},0.5)$. The value of the Fermi vector at
15K was accurately measured by x-ray scattering experiments [5] and found to
be equal to the incommensurate value 2$k_{F}$=0.748$b^{*}$ at 15K. In the
vanadium-doped crystals, previous studies [6] had shown that the low
temperature satellite reflection reduced position $\mathbf{q}_{c}$=($%
1,2k_{F}=0.685,0.5$) was slightly changed with respect to the pure compound.
This value is in accordance with the $2k_{F}$ value expected from the change
in band filling due to the substitution of Mo$^{6+}$ by 2.8\% V$^{5+}$. In
the W-doped crystal, where the tungsten atom is isoelectronic to the
molybdenum, no significant change of the Fermi wave vector was observed.%
\vspace{0.6cm}

\noindent

\begin{figure}[tbp]
\centerline{
\epsfig{file=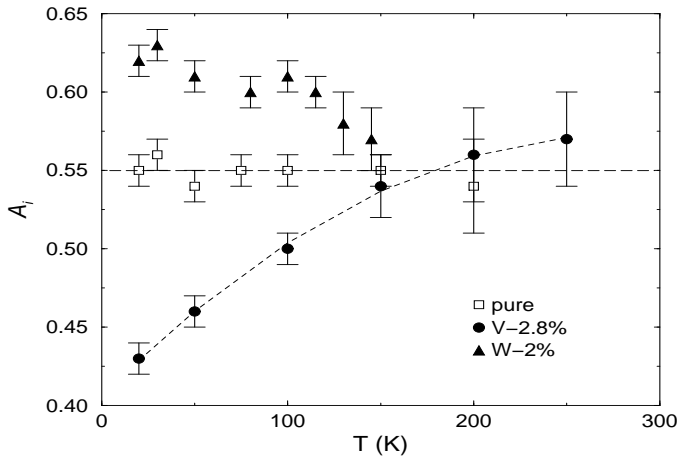,angle=90.0,height=7cm,width=10cm
}}
\caption{Temperature dependence of the asymmetry ratio $A_{i}$ in the pure
(squares), V-doped (circles) and W-doped (triangles) crystals.}
\end{figure}

\vskip .2in \textbf{2. EXPERIMENTAL} \vskip .1in The measurements have been
performed with a three circles diffractometer on a 12kW-rotating anode x-ray
source. High resolution experiment have been performed with a four circles
diffractometer on the D2AM beamline at ESRF, with $\lambda =0.7$ \AA . In
order to investigate the interference effect on the $+2k_{f}/-2k_{f}$
satellite reflections, careful measurements were carried out in the pure,
the V-doped and the W-doped crystals on the same satellite pairs. These
pairs were chosen around the ($\overline{1},7,0.5$) and ($\overline{1},%
\overline{11},0.5$) reciprocal position. The ratio $%
A_{i}=I_{-2k_{F}}/I_{+2k_{F}}$ between the ($\overline{1},\overline{11}%
-2k_{F},0.5$) and ($\overline{1},\overline{11}+2k_{F},0.5$) peak intensity
was measured as a function of temperature in the three compounds. The
temperature dependence of the ratio $A_{i}$ is shown in Figure 1. In the
pure compound, $A_{i}=.55$ remains essentially constant in the temperature
range studied (10 K- 300 K). In the doped crystals however, a clear
deviation from $A_{i}=.55$ is observed below $T\sim $150K. This temperature,
close to $T_{p}$, is the temperature below which the intensity of the
satellite reflection starts to increase in both compounds. In the V-doped
crystals, the value $A_{i}$ is about 20\% lower than in the pure compound
while in the W-doped crystal the value $A_{i}$ is about 10\% larger than in
the pure compound. It is worth noting that the IA occur in opposite
directions in the two types of doped crystals.

In addition, a strong PA has been observed at low temperature in the V-doped
case. Fig. 2 shows the profile of the ($\overline{1},\overline{10.685},0.5$)
satellite reflection in the $b^{*}$ direction at T=60 K. The intensity at
small wave vector (k$\leq $2k$_{f}$) is stronger than the intensity at high
wave vector (k$\geq $2k$_{f}$). Since the peak is much broader than the
experimental resolution, the profile is not affected by spurious resolution
effects. This type of profile was also observed on different satellite
reflections. Due to its unusual shape, we have not attempted to fit the
profile by a special function. In the W-doped crystal, a very slight PA was
observed at low temperature but the profile can be essentially fitted by a
squared Lorentzian lineshape (see fig. 2). At T=15 K, the CDW coherence
lengths $\ell _{b*}$, $\ell _{2a*+c*}$ and $\ell _{2a*-c*}$ at 15K are equal
respectively to 18.5\AA , 7\AA\ and 3.8\AA\ for the V-doped compound [7a]
and 70\AA , 28\AA\ and 7.5\AA\ for the W-doped compound. This shows that the
CDW\ domain of coherence is 2D for the W-2\% crystal whereas it is only 1D
for the V-2.8\% one. There is only one impurity per correlated domain in the
V-doped crystal and about 10 impurities per correlated domain in the W-doped
crystal.\vspace{0.6cm}

\begin{figure}[tbp]
\centerline{
\epsfig{file=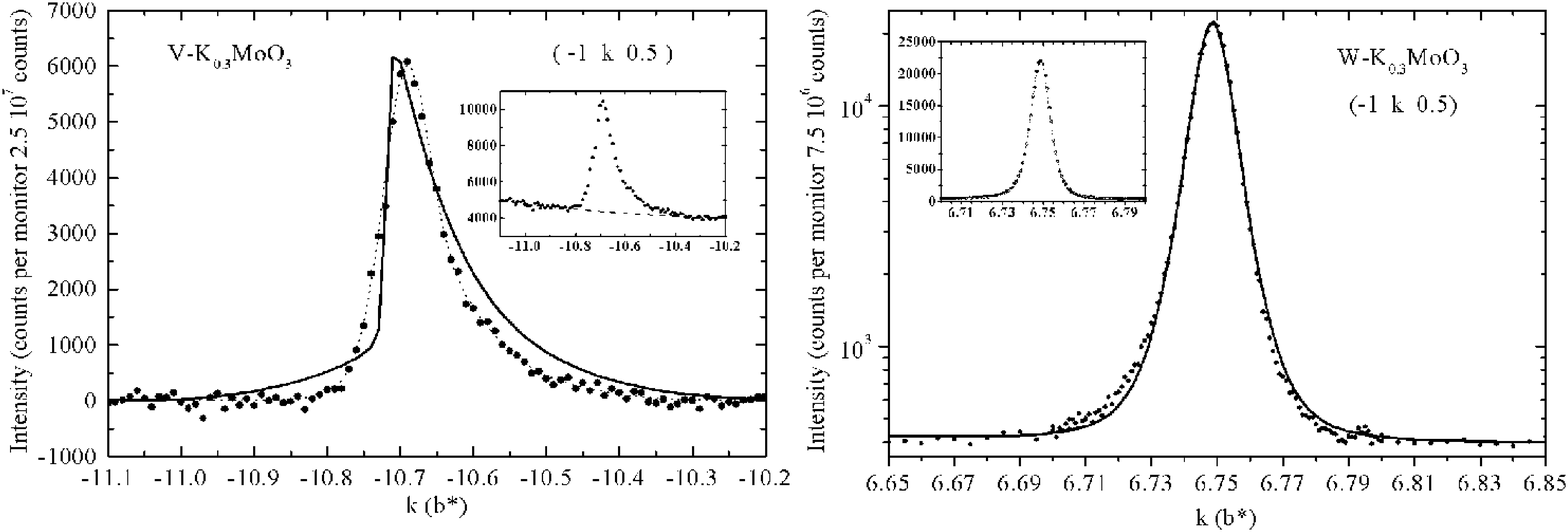,angle=0.0,height=7cm,width=18cm
}}
\caption{Left : Profile of the ($\overline{1},\overline{10.685},0.5$)
satellite reflection in the V-doped blue bronze at 60 K. Solid line is the
intensity obtained from a Friedel oscillation as described in the text.
Right : Profile of the ($\overline{1}, 6.748 ,0.5$) satellite reflection in
the W-doped blue bronze at T=10 K. Solid line is the best fit by a
lorentzian-squared form. Note the logarithmic scale and the normal scale in
the insert.}
\end{figure}

\vskip.2in \textbf{3. RESULTS\ AND\ DISCUSSION } \vskip.1in Let us first
consider the V-doped bronze previously considered in ref. [7]. Due to the
small number of impurities in a coherence domain, one can conclude that the
CDW is strongly pinned to each V-impurity. Using the calculation of ref.
[3], one deduces from the decrease of the $-2k_{F}$ satellite intensity with
respect to the $+2k_{F}$ one, that the neighboring Mo atoms are displaced
away from the vanadium impurity. It is worth noting that this result is
opposite to elastic effects for which molybdenum atoms will move towards the
vanadium atoms of smaller size. Thus this effect clearly originates from an
electronic process. As far as the PA is concerned, its direction indicates
that the $2k_{F}-$CDW expands around the impurity. This corresponds to a
local decrease of the electronic density, needed to screen the supplementary
electron provided by the V$^{5+}$ atom, with respect to the Mo$^{6+}$ atoms
[7b]. The case of the W-doped crystals is more subtle. The correlated
domains contain 10 impurities on average, which is more consistent with a
weak pinning scenario. Nevertheless, the observed IA indicates that the
phase of the CDW adjusts as a whole on the impurities positions in a
coherence domain. This global adjustment is expected in a weak pinning case.
Furthermore, the Lorentzian squared profile of the satellite reflection is
also expected in the weak pinning situation.

Let us now discuss in more detail the microscopic interactions between the
CDW and the impurity potentials and the consequence of such interactions on
the x-ray scattering intensity. At low temperature, when the CDW is well
developed the Hamiltonian $\mathcal{H}$ describing the interaction of the
CDW phase with the impurity reads :

\begin{eqnarray}
\mathcal{H} &=&\frac{1}{s}\int d\vec{r}\{C_{\parallel }/2\left( \partial
_{x}\varphi \right) ^{2}+C_{\bot }/2\left( \partial _{\bot }\varphi \right)
^{2}+\sum_{m}V(\vec{r}\mathbf{-}\vec{r}_{m})\partial _{x}\varphi /\pi  \\
&&-U\cos [q_{c}\vec{r}+\varphi (\vec{r})]\,\,\delta (\vec{r}-\vec{r}_{m})\},
\end{eqnarray}
where $s$ is an area per chain and $C_{\parallel }$ and $C_{\bot }$ are the
elastic moduli of the CDW in the chain direction and normal to the chains
respectively. $\varphi =\varphi (\vec{r})$ is the CDW phase and $\vec{r}_{m}$
are the positions of the impurities. The impurity potential has been
separated into a long range part $V$, leading to the forward scattering of
the electron in the metallic state, and a short range part $U$, leading to
its backward scattering. More generally, $V$ is the potential interacting
with the non oscillating part of the charge density $\pi ^{-1}\partial
_{x}\varphi $. From the previous equation, one can see that the effect of
the backward scattering potential $U$ is to pin the phase of the CDW at the
impurity position, while the effect of $V$ is to stretch/compress the phase
of the CDW. As a general result, the x-ray IA comes from the backward
scattering potential $U$, while the PA originates from the forward
scattering potential $V$.

Let us now consider the effect of Friedel oscillations. It is well known
that in a metal, the screening of an impurity charge occurs via the phase
shift $\eta $ of the electronic wave function scattered by its potential
[8]. In 1D, the resulting charge oscillation $\delta \rho $ is coupled to
the distortion $u(x)$, which thus reads at large $\left| x\right|$ 
\begin{equation}
u(x)\simeq \frac{1}{\left| x\right| }\sin (2k_{F}x+\varphi (x)),
\end{equation}
where $\varphi (x)$ is the FO phase so that $\varphi (\infty )=\eta $ and $%
\varphi (-\infty )=-\eta $. The phase shift $\eta $ is related to the charge 
$Z$ of the impurity by the Friedel sum rule :

\begin{equation}
Z=\frac{2}{\pi }\eta .
\end{equation}
In the case of a single charged vanadium impurity one has $\eta =\pi /2$.
This means that the complete screening is achieved by stretching the CDW
exactly by $\pi $. The total phase shift $\delta \varphi =\pi $ of the CDW
has to be accumulated on a very short distance $d$. Figure 2a presents the
experimental results together with the computed intensity of the scattering
by a FO described by the equation (2) and $d\sim $1 \AA . As expected from
more general considerations on the x-ray scattering [3], this computed
profile is essentially asymmetric and bears a striking resemblance with the
experimental results. A complete discussion of the scattering by a FO, taken
into account the damping of the FO by the microscopic coherence length $\xi
_{0}$ of the electrons will be presented elsewhere The main conclusion of
this study is the evidence of FO in the vicinity of the vanadium atoms in
2.8\% V-doped blue bronze. IA have also been observed, which provides
evidence of a coherence between the impurity position and the CDW/FO. The
W-doped crystals are more likely described by a weak pinning scenario, in
which the phase of a correlated domain adjusts as a whole. Nevertheless, the
observation of a PA for all the satellite reflections studied clearly
indicates the presence of phase distortions, corresponding to a local
decrease of the electronic density. Additional experiments on crystals with
different doping levels are planned in order to elucidate more
quantitatively the microscopic features of the pinning of CDWs in low
dimensional materials.

\noindent \textbf{Acknowledgments } We thank J.F. B\'{e}rar and R. Moret for
their help during the synchrotron experiments.\vspace{1cm}

\noindent \textbf{References} \vspace{0.5cm}

\noindent 1. See e.g. G. Gr\"{u}ner, \textit{Density Waves in Solids} ,
(Frontiers of Physics Series, Addison-Wesley, Massachusetts, 1994)\newline
2. I. T\"{u}tt\'{o} and A. Zawadowski, Phys. Rev. B\textbf{32}, (1985) 2449;
J. R. Tucker, Phys. Rev. B\textbf{40}, (1989) 5447.\newline
3. S. Ravy, J.-P. Pouget and R. Com\`{e}s, J. Phys. I (France) \textbf{2},
(1992) 1173; S. Ravy, J.-P. Pouget, J. Phys. IV (France) \textbf{3}, (1993)
109.\newline
4. S. Brazovskii, J.-P. Pouget, S. Ravy and S. Rouzi\`{e}re, Phys. Rev. B%
\textbf{55}, (1997) 3426. \newline
5. J.-P Pouget, in \textit{Low-dimensional electronic properties of
Molybdenum Bronzes and Oxides, }Ed. by C. Schlenker (1989), Kluwer Academic
Publisher.\newline
6. S. Girault, A.H. Moudden, J.-P. Pouget and J.-M. Godard, Phys. Rev. B%
\textbf{38}, (1988) 7980.\newline
7a) S. Rouzi\`{e}re, S. Ravy, J.-P. Pouget, Synth. Met. \textbf{70}, 1259
(1995);b) \textbf{90}, (1997) 2131.\newline
8. J. Friedel, Phil. Mag. \textbf{43}, 153 (1952).\newline
9. S. Rouzi\`{e}re, S. Ravy, S. Brazovskii and J.-P. Pouget, in preparation.

\end{document}